\documentclass[12pt,a4paper]{JHEP3}
\usepackage{latexsym,graphicx,epsfig}

\newcommand{\be}{\begin{equation}}
\newcommand{\ee}{\end{equation}}

\newcommand{\bea}{\begin{eqnarray}}
\newcommand{\eea}{\end{eqnarray}}

\title{{Non-equilibrium Higgs transition} \\
{in classical scalar electrodynamics}
} 
\author{{D\'enes Sexty and Andr\'as Patk{\'o}s}\\
{Department of Atomic Physics}\\
{E{\"o}tv{\"o}s University, Budapest, Hungary}\\
\it E-mail: \email{denes@achilles.elte.hu} and
\email {patkos@ludens.elte.hu}
}


\abstract{
Real time rearrangement of particle spectra is studied numerically in
a $U(1)$ Gauge+Higgs system, in the unitary gauge and in three spatial
dimensions. The cold system starts
from the symmetric phase. Evolution of the partial energy densities
and pressures reveals well-defined equations of state for the
longitudinal and transversal gauge fields very early. Longitudinal
modes are excited more efficiently and thermalize the
slowest. Hausdorff-dimension of the Higgs-defect manifold, eventually seeding
vortex excitations is thoroughly discussed. Scaling dependence of
the vortex density on the characteristic time of the symmetry breaking
transition is established. 
}

\begin{document}

\section{Introduction}
In hybrid models of inflation \cite{linde94}
a non-equilibrium Higgs transition leads to
the (p)reheating of the universe. The accompanying spinodal instability  
\cite{felder01,asaka00} determines the initial particle composition
and also might have important impact on baryogenesis \cite{smit04}. 
The emerging particle composition and thermalised collective behavior  
of the fields represent starting data for the Hot Universe.
   
The process of field excitation was thoroughly studied by Skullerud  
{\it et al.} in the classical $SU(2)$ Higgs model as it occurs 
after a step function sign change (quench) in the squared mass
parameter of the Higgs potential \cite{skullerud03}. 
 Particle numbers and particle energies were extracted from
 correlation functions in analogy with free field theories.  The
 quench leads both in
Coulomb and unitary gauges to relativistic dispersion relations
 and well defined effective masses after a certain 
characteristc time, which is much shorter than what is required for reaching
classical thermal equilibrium. In the unitary gauge, important 
differences were observed between the degree of excitation
 of the transverse and longitudinal gauge polarisations.

In a previous paper \cite{sexty05} we proposed methods for
 investigating the partial
pressures and energy densities associated with the Higgs field, and the
longitudinal and transversal parts of the gauge fields of scalar
electrodynamics in thermal equilibrium. These quantities were defined
in the unitary gauge by splitting the diagonal elements of the
energy-momentum tensor of the system into three pieces, 
following the intuition gained with constant Higgs background. 
It was shown that all three pieces obey separately 
quasi-particle thermodynamics in the broken phase. In particular, 
with help of the  concept of spectral equations of
state effective masses were extracted  
for all three quasi-particles in agreement with the expected
degeneracy pattern and magnitude. It is worth to mention that in this
analysis the explicit construction of the quasi-particle coordinates
could be avoided. It is appealing to extend this thermodynamical
approach also to the non-equilibrium evolution relevant to
the post-inflationary scenario (see \cite{podolsky05}). 
This is the first subject to be discussed in this paper (section 3).

In a parallel direction of research Hindmarsh and Rajantie 
\cite{hindmarsh00,hindmarsh01} studied the vortex generation  
in $U(1)$ gauge+Higgs systems when a change of sign occurs in the squared mass
parameter of the theory gradually with time scale $\tau$. The
frequency of vortex generation scales with some power of the quenching time. 
They derived the corresponding scaling laws from the proposition
 that the dominant mechanism of vortex 
formation is the trapping and smoothing of the fluctuating magnetic flux of the
high $T$ phase into vortices. It is the ordered magnetic flux
 which induces linear defect formation in the Higgs field, which locally
 minimizes the energy density of the system. 

The scenario may differ when the
temperature of the starting system is zero, and magnetic
fluctuations actually build up in the excitation process  after the
quench, which is the case at the end of inflation. There
is some chance that short length linear pieces of Higgs defects, where 
the field stays near zero after the average has rolled down from the top of the
potential to its
symmetry breaking value, join each other. Coherent
excitation of the surrounding magnetic flux might stabilize the defect 
line in form of Nielsen-Olesen vortices.

The first stage of this latter scenario is the original 
Kibble-Zurek defect formation \cite{kibble76,zurek85}. It implies 
a unique early time variation of the defect densities in all
configurations independently whether all defects  dissipate in a few
oscillations of the average order parameter 
or (quasi)stable vortices are ''successfully'' formed. 
 The  manifold of the Higgs-zeros might  
  contain at early times various objects, e.g. isolated point defects,
  two-dimensional domain walls or blobs of some fractal
  dimensionality in addition to the sites which will be eventually
 associated with  topologically stabilized 
vortices. In order to focus the study on the early
 vortex statistics one has to find the earliest possible
  time interval where this complex manifold is dominated by nearly
  one-dimensional objects.
In the present study we propose for this purpose to measure the
  time dependence of
 the Hausdorff-dimension ($d_H$)
of the manifold of Higgs-zeros in a broad range of the model parameters,
including possible dependence on the discretisation. We have checked
that an extensive domain of the parameters exists where this dimension
is close to unity in an extended time interval. 
We have tested that the sustained $d_H\approx 1$ ensures 
that the  objects emerging after longer time evolution
predominantly form linear (stringlike) excitations. 
The roll-down time ($\tau_r$) of the order parameter was also introduced,
measuring the time elapsed until the order parameter starting
  from the unstable symmetric extremum passes the first
time its maximum (which overshoots the  
symmetry breaking equilibrium value).
The density of the vortex excitations shows powerlike dependence  on $\tau_r$, 
when runs are compared for which the energy density is kept at 
some fixed value, while the location of the symmetry breaking minimum
is varied.  This is the second issue to be discussed in 
this paper (section 4).

\section{Set-up of the numerical study and sample selection}

Time evolution of the classical $U(1)$ Higgs model 
is tracked by solving the corresponding equations of motion
\bea
&
\displaystyle
D_\mu D^\mu\Phi({\bf x},t)+m^2\Phi({\bf
  x},t)+\frac{\lambda}{6}|\Phi({\bf x},t)|^2\Phi({\bf x},t)=0\nonumber\\
&
\displaystyle
\partial_\mu F^{\mu\nu}+\frac{ie}{2}(\Phi^*D^\nu\Phi-\Phi
D^\nu\Phi^*)=0,
\eea
where $D^\nu=\partial^\nu+ieA^\nu$ is the covariant derivative defined
with the vector potential $A^\nu$, and $F^{\mu\nu}$ is the Abelian
field strength tensor. The equations were initialized and solved in 
the $A_0=0$ gauge
and the solution was transformed to the unitary gauge where the
degrees of freedom appear the closest to their expected physical
multiplet structure in the Higgs phase. 

The system starts from a symmetric $(\Phi =0)$ initial
state. The inhomogenous scalar modes of the complex Higgs field 
fluctuate initially with a phase and amplitude distribution 
corresponding to the  stable symmetric (e.g. $m^2 > 0$) $T=0$
  vacuum. The fluctuations have to respect 
the global neutrality constraint of the system, which was imposed on the 
Monte Carlo sampling of the initial configurations. 

 At $t=0$ an instant quench, $m^2\rightarrow -m^2$, is
  performed. Then the Fourier modes ${\bf k}^2 < |m^2|$ of the scalar
  field start an
   exponential growth in time due to the spinodal
  (tachyonic) instability. The higher $k$ scalar modes and the gauge
  fields will be excited only afterwards by effective source terms
  built overwhelmingly from highly excited spinodal modes. 
Therefore the evolution of
  the system will be not sensitive to the accurate initialisation of the
  non-spinodal modes, if the
  contribution to the energy density from these modes is negligible
  relative to the gap in the potential energy density. The simplest is
to choose the amplitude of the non-spinodal scalar and all gauge modes
to be zero,  except the longitudinal field strength, 
$ {\bf \Pi}_L \equiv {\bf E}_L $, which was calculated from the Gauss 
constraint: $ \nabla {\bf \Pi}_L = e^2 \textrm{Im} \Phi^* \dot\Phi $.

 The spectra of the initial scalar excitations is cut at $|m|$, 
therefore all
observables are UV-finite. Until the highest Fourier modes are not
excited these quantities remain insensitive to the lattice spacing
 and the results can be
interpreted without any need for lattice spacing dependent
renormalisation. \cite{skullerud03,smit01}

The equations were  discretized
in space and time. The independent parameters of the discretized
system  are the following: $ dx|m|, dt|m|, \lambda$ and $e^2$. The
ratio $dt/dx$ was kept fixed ($\sim 0.1$) as well as the gauge
coupling $e=1$. The investigation concerning the emergence of the Higgs
equation of state during the transition (section 3) 
was realised for $dx|m|=0.35,\lambda=6$. The Hausdorff-dimension of 
the Higgs-defect manifold
was determined for an extended domain of the ($dx|m|,\lambda$)-plane 
(see section 4).  Lattices of size $ 64^3, 96^3, 128^3$ were studied.

The runs could be divided easily
 into classes according to the (quasi)stationary
field correlations  emerging after the fastest transients are relaxed.
These correlation coefficients are defined as
\be
\Delta[\Psi_1,\Psi_2](t)=\frac{\overline{\Psi_1^2\Psi_2^2}
-\overline{\Psi_1^2}~\overline{\Psi_2^2}}{\overline{\Psi_1^2\Psi_2^2}}
\ee
for the fields $\Psi_1, \Psi_2$, with the overlines denoting spatial
averages at time $t$.
Nonzero (quasi)stationary values of $\Delta [{\bf A}_T,\rho], \Delta [{\bf
  B},\rho]$ in the unitary gauge (where $\rho\equiv|\Phi|$, 
${\bf A}_T$ denotes the transverse
part of the vector potential, and $\bf B$ is the magnetic field strength) 
perfectly signal  the presence of vortex-antivortex pairs.
In equilibrium (vortex-less) configurations these coefficients take
values compatible within the fluctuation errors with zero, as shown in
our earlier paper \cite{sexty05}. The emergence of the equations of
state for the different quasiparticle constituents of the system was
studied exclusively in vortex-free configurations. 

\section{Emerging equations of state}

The expressions of the constituting energy
densities and partial pressures in the unitary gauge are the following:
\bea
\epsilon&=&\epsilon_\rho+\epsilon_T+\epsilon_L,\qquad
p=p_\rho+p_T+p_L,\nonumber\\
\epsilon_\rho&=&\frac{1}{2}\Pi_\rho^2+\frac{1}{2}(\nabla\rho)^2+V(\rho),
\qquad p_\rho
=\frac{1}{2}\Pi_\rho^2-\frac{1}{6}(\nabla\rho)^2-V(\rho),
\nonumber\\ 
\epsilon_T&=&\frac{1}{2}[{\bf \Pi}_T^2+(\nabla\times{\bf
    A}_T)^2+e^2\rho^2{\bf A}_T^2],\qquad
p_T=\frac{1}{6}[{\bf \Pi}_T^2+(\nabla\times{\bf
    A}_T)^2-e^2\rho^2{\bf A}_T^2],\nonumber\\
\epsilon_L&=&\frac{1}{2}\left[{\bf \Pi}_L^2+e^2\rho^2\left({\bf
    A}_L^2+\frac{1}{(e^2\rho^2)^2}(\nabla\Pi_L)^2\right)\right],
\nonumber \\
p_L&=&\frac{1}{6}[{\bf \Pi}_L^2-e^2\rho^2{\bf A}_L^2]+
\frac{1}{2}\frac{1}{e^2\rho^2}(\nabla{\bf\Pi}_L)^2.
\label{pressures}
\eea
The field $A_0$ has been eliminated with the Gauss-constraint.

In the process of the excitation initiated by
 the instability the constituting energy densities and
 pressures vary in time. Mapping the trajectory of the three species
in the $(\epsilon - p)$-plane one can determine the time scale needed for
establishing linear relations ($p=w\epsilon$)
characteristic for the equation of state of a nearly ideal gas
in equilibrium. In case of the sample not containing
vortices such simple trajectory appears for the gauge degrees of
freedom fairly early with $w > 0$,  as can be seen from
Fig.\ref{Fig:1}. One observes that the early (lower energy density)
portion of the $p-\epsilon$ trajectory of the transversal gauge field
is somewhat steeper than its later (higher energy density) piece.
This deviation signals the presence
of quickly evaporating small size vortex rings. The defect network in
 these configurations evaporates by the time $ (40-50)|m|^{-1}$.

\TABLE{
\label{Tab:slopes}
\begin{tabular}{|l|c|c|c|}
\hline
&$L=64$&$L=96$& $L=128$\\\hline
H&-0.16$\pm$ 0.23& -0.10$\pm$ 0.05&-0.09$\pm$ 0.02 
\\\hline
T&0.18$\pm$ 0.05& 0.18$\pm$ 0.03& 0.16$\pm$ 0.02
\\\hline
L&0.13$\pm$ 0.04& 0.15$\pm$ 0.03& 0.12$\pm$ 0.05
\\\hline
&$n=70$& $n=82$ & $n=17$
\\\hline
\end{tabular}
\caption{Size dependence of the slope parameter $w=p/\epsilon$ 
of the equations of
  state of the different particle species (H= Higgs, T= transverse, L=
  longitudinal). $L$ gives the size of the lattice, $n$ is the number
  of simulations included into the statistics}
}

In Table 1 the size dependence of the average slope $w$ appears for the three
field excitations as measured on different lattices. 
The average slopes for the vector and the longitudinal mo\-des are
compatible. The negative slo\-pe found for the Higgs field is
interpreted very naturally. 
The energy was stored fully in the low-$k$ (spinodal) modes of this
component directly after the instability was over. 
It was transferred in the later evolution to the transverse gauge fields
at about the same rate as to the high-$k$ Higgs modes. The latter
positively contribute to the full Higgs pressure at the same time
when the Higgs field globally looses energy. In this way one naturally arrives
to a trajectory with negative slope. 
This scenario can be tested relatively easily: displaying 
only the partial pressure and energy density due to the 
high-$k$ Higgs modes, their EOS should be closer to the ideal 
radiation equation of state ($ w =1/3 $), that is the equation of
state of these modes should show positive slope similarly as the gauge
fields do. This trajectory appears in the left hand end of
Fig.\ref{Fig:1}. This curve is actually steeper than the $p-\epsilon$
line of the massless limiting case, which is a clear signal of
non-equilibrium.

\FIGURE{
\includegraphics*[width=12cm]{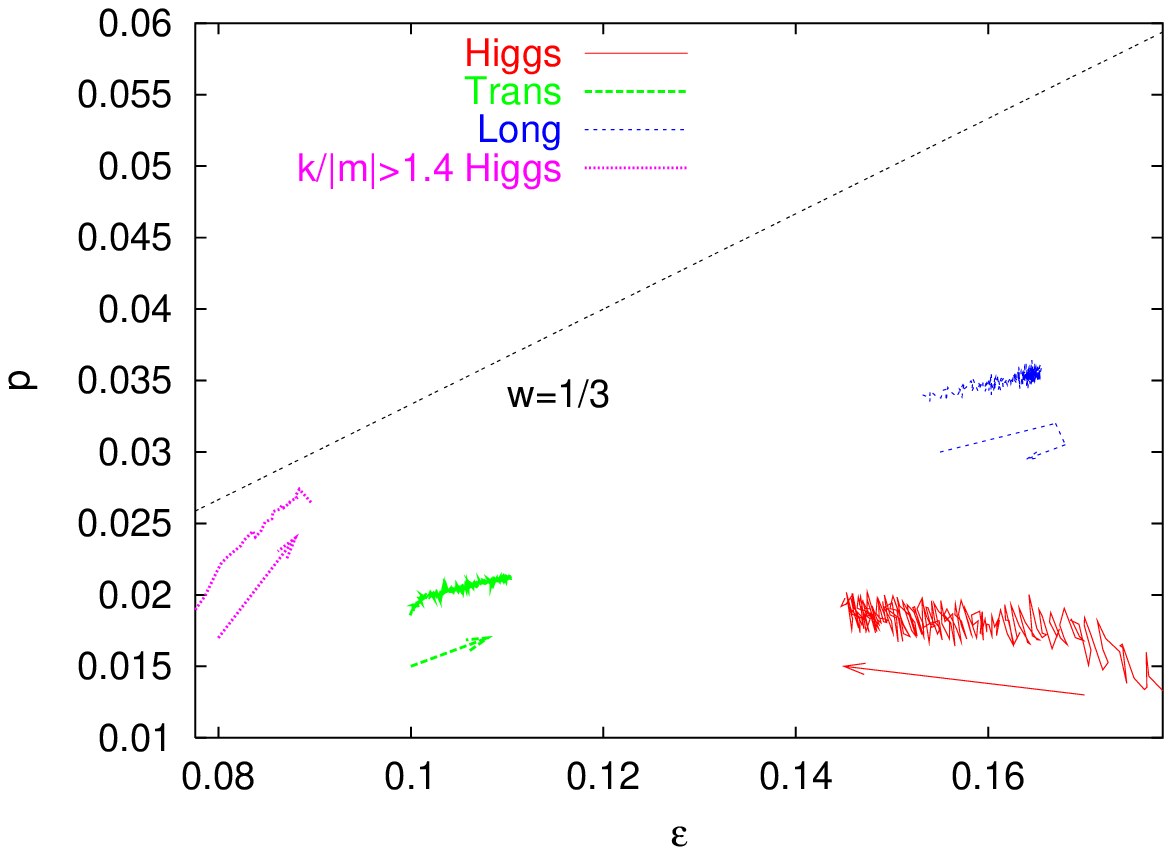} 
\caption{Trajectories of the transversal and longitudinal gauge
 and the scalar Higgs degrees of freedom in the energy-density - pressure
 plane (both measured in $|m|^4$ units). 
The longitudinal degree of freedom receives 60-70\% more energy
 from the spinodal instability than the transversal
polarisations. The gauge fields display linear equation of state very
 early. The
average slope of the path of the Higgs field is negative. Restricting
 to the contribution of high $k$ modes the trajectory will have
 positive slope (cf. left edge of the figure). The arrows show the
 time direction.}
\label{Fig:1}
}

Another aspect of the degeneracy of the longitudinal and transversal
gauge degrees of freedom can be observed when studying the so-called spectral
equations of state, introduced in our earlier investigation
\cite{sexty05}.  Using the spatial Fourier transform of the square-root
of the pressure and of the energy density distributions
one can compute a spectral power for these quantities.
Without any need for explicit construction of the quasi-particle
coordinates one implicitly assumes that near
equilibrium each mode ${\bf k}$ can be described by small amplitude
oscillations of some effective field coordinate and its conjugate
momentum. The average kinetic and potential energies of a harmonic
oscillator are equal. With this assumption one finds:
\bea
&
\displaystyle
\overline{\epsilon_\rho({\bf k})}={\bf k}^2\overline{\rho^2}+
2\overline{V(\rho)},\qquad
\overline{p_\rho}=\frac{1}{3}{\bf k}^2\overline{\rho^2},\nonumber\\
&
\displaystyle
\overline{\epsilon_T({\bf k})}={\bf k}^2\overline{{\bf A}_T^2}+
e^2\overline{\rho^2{\bf A}_T^2},\qquad \overline{p_T}=\frac{1}{3}{\bf
  k}^2\overline{{\bf A}_T^2},\nonumber\\
&
\displaystyle
\overline{e^2\rho^2\epsilon_L}={\bf k}^2\overline{\Pi_L^2}+
\overline{e^2\rho^2\Pi_L^2} , \qquad \overline{e^2\rho^2 p_L}=\frac{1}{3}
{\bf k}^2\overline{\Pi_L^2}.
\eea
If the effective squared mass combinations 
\be
M^2_{eff,\rho}\equiv 2\frac{\overline{V(\rho)}}{\overline{\rho^2}},
\qquad
M^2_{eff,T}\equiv e^2\frac{\overline{\rho^2{\bf
    A}_T^2}}{\overline{{\bf A}_T^2}},\qquad
M_{eff,L}^2\equiv
e^2\frac{\overline{\rho^2\Pi_L^2}}{\overline{\Pi_L^2}}
\ee
do not vary with $\bf k$ then the following generic behavior can be 
seen for the spectral equations of state:
\be
w_i({\bf k})=\frac{\overline{p_i({\bf k})}}{\overline{\epsilon_i({\bf
      k})}}=
\frac{1}{3}\frac{{\bf k}^2}{{\bf k}^2+M_{eff,i}^2}.
\label{spect-eq}
\ee
In the twin figures of Fig.\ref{Fig:2} one can gain impression of how the
spectral equation of state develops in time for the different
species. 
The transversal mode displays a behavior which follows
(\ref{spect-eq}) almost perfectly extremely early. 
Its mass can be extracted from the fit very reliably. 
Although the high $k$ part of the longitudinal energy density stays anomalously
large and therefore the corresponding spectral equation of state drops
to zero above some $k$ value, in the low $k$ region its degeneracy
with the transversal polarisation
(signalling the same mass value) is well fulfilled for $|m|t\geq 60$.

\FIGURE{
\includegraphics*[width=7cm]{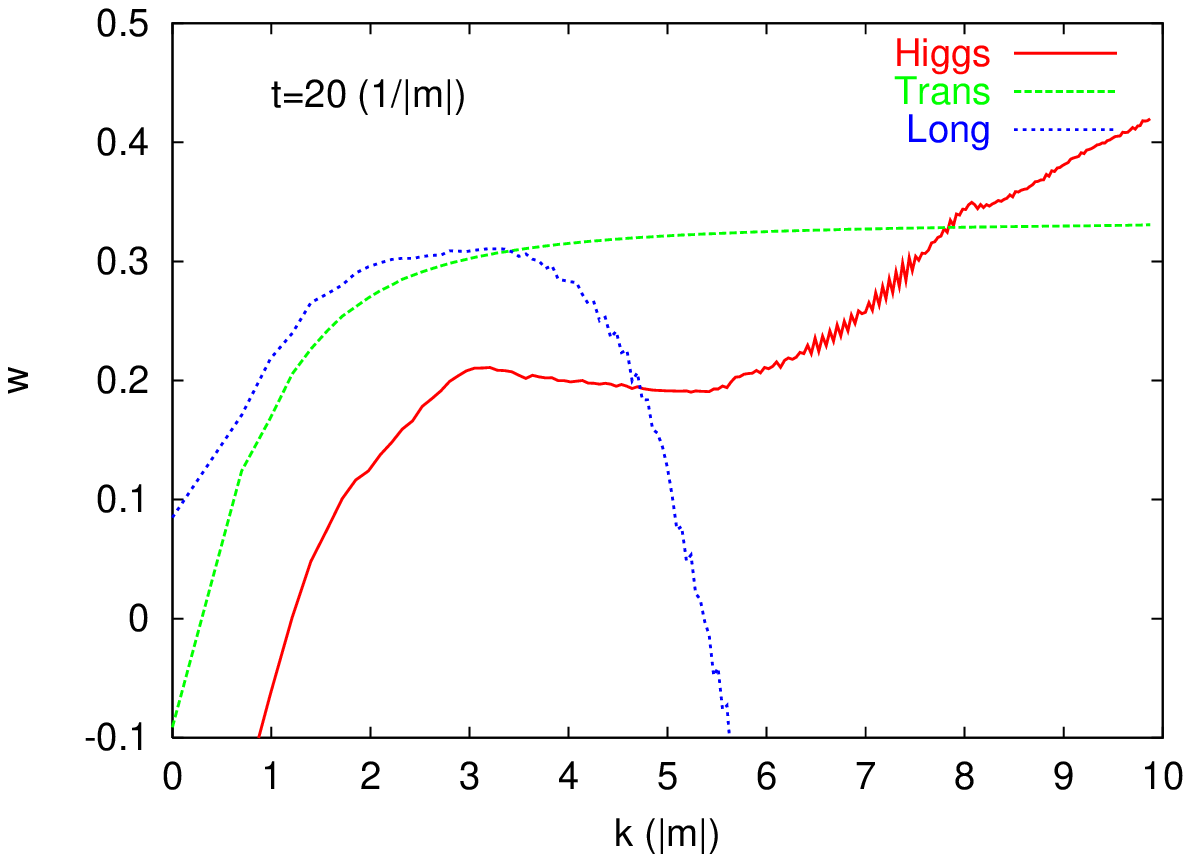} 
\includegraphics*[width=7cm]{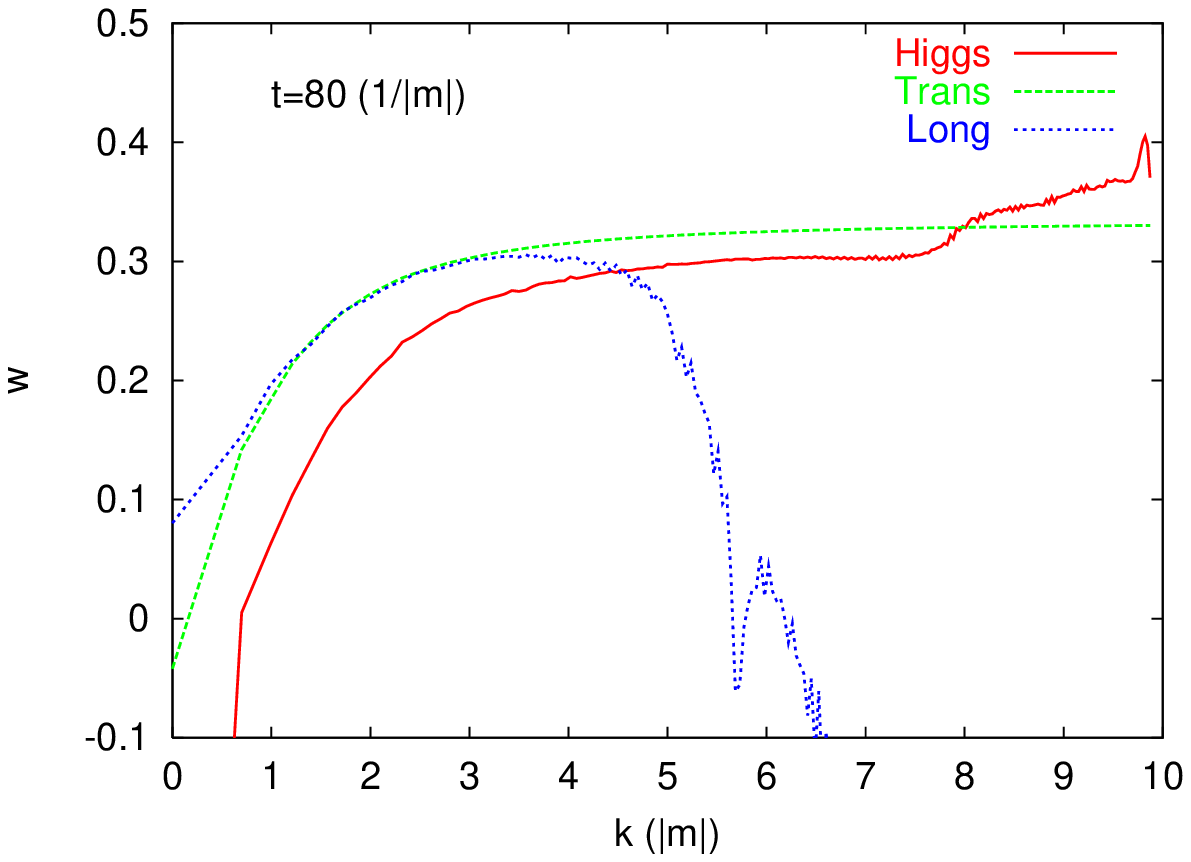}
\caption{The process of establishing the spectral equation of state
  for the three particle species. The curves on the left figure
  are plotted about the time when the linear regime of the global
  equations of state sets in. Important deviation of the measured curve from
  (\ref{spect-eq}) 
is observed for the Higgs field and also for the longitudinal 
gauge polarisation. The agreement improves substantially 
at later times (figure on the right).  The more persistent deviations 
in the low $k$ region signal a slow equilibration. ($N=96$, average of 20 
independent runs.)
}
\label{Fig:2}
}

\FIGURE{
\includegraphics*[width=11cm]{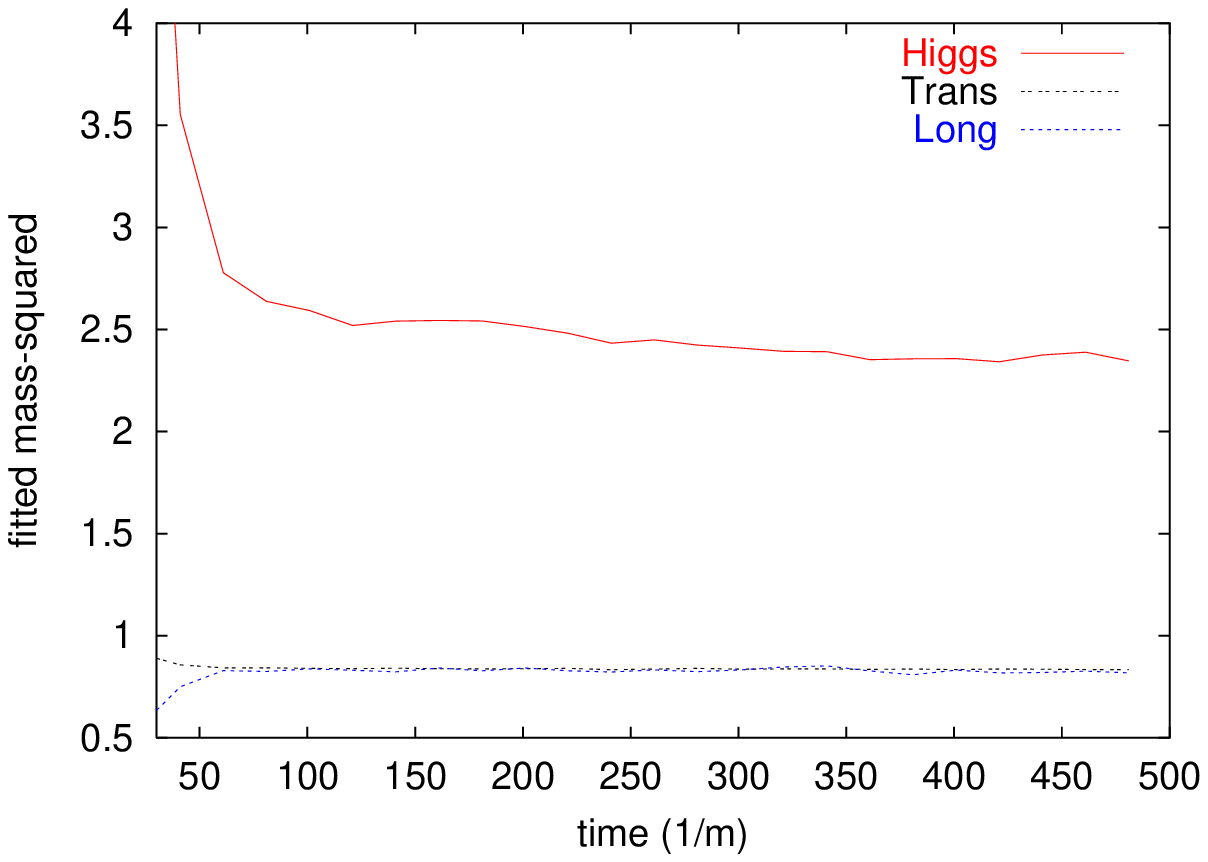} 
\caption{ Evolution of the mass-squared values fitted to 
the spectral EOS with (\ref{spect-eq}). The fitting region 
was: $0.3 < k  < 5 $. The error of the fitted values is 
approximately 5\%. The gauge modes show degeneracy after 
around $t = 60/|m|$.
}
\label{Fig:mass}
}

The onset of the Higgs effect is quantitatively characterised
by the degeneracy of the masses fitting the spectral equations of both
gauge polarisation states (Fig.\ref{Fig:mass}).
 At the earliest
times the longitudinal state is considerably lighter, but we were not
able to find an intrinsic Goldstone behavior even at the earliest
times when
the spectral equation of state already worked.

It is surprising that the spectral equation of state of the Higgs
field has qualitatively the expected form in spite the fact that it is
far from equilibrium. Its mass value slowly (nearly linearly) relaxes to
the non-renormalised (lattice spacing dependent) mass value measured
in \cite{sexty05}.
 This suggests that the mode-by-mode equality of the
average kinetic and potential energies is reached very quickly, just
the distribution of the power 
among the different modes is far out of equilibrium.

We conclude this section by discussing
 some global features of the degree of excitation of the
different particle species and their relaxation towards equilibrium.
 In equilibrium configurations we have
shown earlier that the share of the three pieces in the energy density
follows the expected 1:1:2 proportion for Higgs, longitudinal gauge and
transversal gauge degrees of freedom, respectively.

The main qualitative feature of our simulation is that the
longitudinal vector mode is excited during the spinodal instability
much more efficiently than the transversal
gauge modes. The energy density does not seem to depend qualitatively
on whether there are vortices present. The energy exchange on the other
hand is much more efficient between the Higgs and the transversal vector modes,
than with the longitudinal part. This behavior is very reminiscent of
the very slow relaxation of the Goldstone modes in systems which go
through the breakdown of a global symmetry \cite{tytgat96,borsanyi02}.

In the case of the vortexed sample an upward deviation can be observed on
both
the Higgs and the transverse gauge trajectories in the $(p,\epsilon)$ plane 
towards higher pressures during the annihilation. The sum of
the transversal and the Higgs energy densities stays nearly constant during 
annihilation, while their pressures shoot up. 
This is intuitively expected since the contribution from the vortex
solution to the pressure is negative. The
annihilations speed up considerably the energy exchange between
these degrees of freedom. The
longitudinal modes hardly participate in this process. 
In an expanding universe one might therefore conjecture that
the longitudinally polarised gauge particles never reach thermal
equilibrium and when decouple they might store much more energy than
one would deduce from calculations based on the assumption of thermal 
equilibrium.

\section{Hausdorff-dimension of the Higgs-defect manifold}

In the $U(1)$ Higgs system there are stable topological defects:
the Nielsen-Olesen vortices, which are identified as
 one dimensional defect lines in the Higgs-field.
Spontanously emerging long-lived vortex-antivortex configurations,
which wind around the whole lattice,
 thus can be characterized by a stationary non-zero
 volume fraction of near-zero sites
$(\rho< \rho_{th})$ of the Higgs field. The volume fraction grows with
 the increase of the threshold value $\rho_{th}$. 

Vortexed configurations branch-off at a certain time from the 
uniform exponentially decreasing tendency of the volume fraction of
low Higgs values, characteristic for all runs at early times.  
The branching happens at around $ t=50/|m| $. Until this happens,
 the fraction of the low Higgs values already performed
several oscillations. This makes clear that building up of stationary vortex
configurations requires considerable time beyond the first appearence
of islands of near zero Higgs values after the first roll-down. 
  Not all near zero sites necessarily belong to some vortex. Lower
  and higher dimensional (possibly) fractal submanifolds will decay
  and should be left out of consideration when one asks for the
 early time density of those Higgs-defects which eventually end up in
  quasilinear, long lived vortex configurations.
A sustained non-zero value 
of the volume fraction of such defects
indicates the presence of a (quasi) 
stationary vortex system until a sudden jump to near-zero values
signals the annihilation of the vortex network. 
	
Our final goal is to analyse the variation of the
density of the emerging one-dimensional
Higgs defect manifold when some characteristic time of
the nonequilibrium transition is varied. Earlier investigations
\cite{copeland02} were mainly done in two dimensions and identified
visually the corresponding topological defects
(domain walls). Deviations from the predicted scaling law for the summary
length of the domain walls were associated with the contamination arising 
from random sign-changes of the field, which were abundantly present in the
early time evolution of
the system. The systematic selection of objects with uniform dimensionality
is an unavoidable precondition since any Kibble--Zurek-type
analysis makes sense only for objects of well-defined dimensional
extension \cite{laguna97,laguna98}. For the Abelian Higgs model the
late(!) time evolution of well developed three-dimensional
vortex networks was followed 
 by identifying the vortices through gauge invariant zeros of
the Higgs field and also by identifying bits of a vortex through $2\pi$
winding of the $U(1)$ phase \cite{vincent98}. 

 The study of the statistics of the vortex generation requires the
identification of Higgs defect lines the earliest
possible time after the spinodal instability starts.
Here we propose to introduce  a ''filtering'' step into the
identification process of strings: the
measurement of the Hausdorff-dimension ($d_H$) of the
defect manifold consisting of near zero values of the Higgs field.
We shall see that it changes during the roll-down of the system and
will be able to find systematically the time interval and the range of
the coupling parameters where mainly one-dimensional objects are present. 

Let us perform a sequence of blocking transformations which 
leads to the determination of $d_H$.
One starts by defining the lattice site manifold of the defects for
the original Higgs-configuration $\rho_{old}({\bf x})$:
\be
X_{old}[\rho_{th}]=\{ {\bf x}=(l,m,n) | \rho_{old}(l,m,n)<\rho_{th} \}.
\ee 
For the formation of the new manifold
 one considers the blocked lattice with rescaled lattice spacing
 $pdx$ and site coordinates $(L,M,N)p
dx$. The value of the block field is defined as
\be
\rho_{new}(L,M,N)=\min \{\rho_{old}(l,m,n)| l=Lp+i, m=Mp+j, n=Np+k,
0\leq (i,j,k) < p\}.
\ee
The blocked defect manifold is defined as 
\be
X_{new}[\rho_{th}]=\{ {\bf x}=(L,M,N)|\rho_{new}(L,M,N)<\rho_{th} \}.
\ee
The number of blocks belonging to $X$ 
should scale with the dimension of the manifold if it corresponds to
objects of well-defined dimensionality embedded in the
three-dimensional space:
\be
\frac{N(X_{new})}{N(X_{old})} = p ^{-d_H}.
\label{hausdorf}
\ee

\FIGURE{
\includegraphics*[width=7cm]{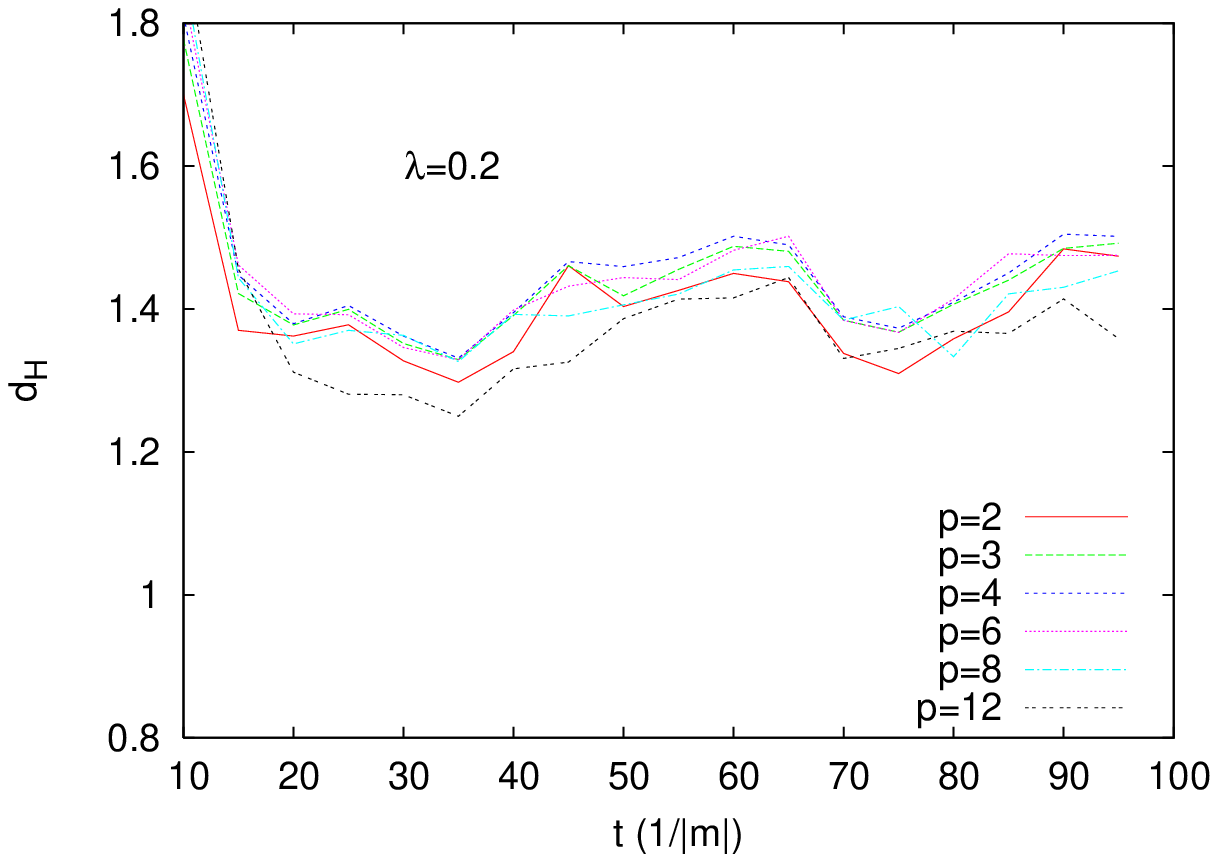} 
\includegraphics*[width=7cm]{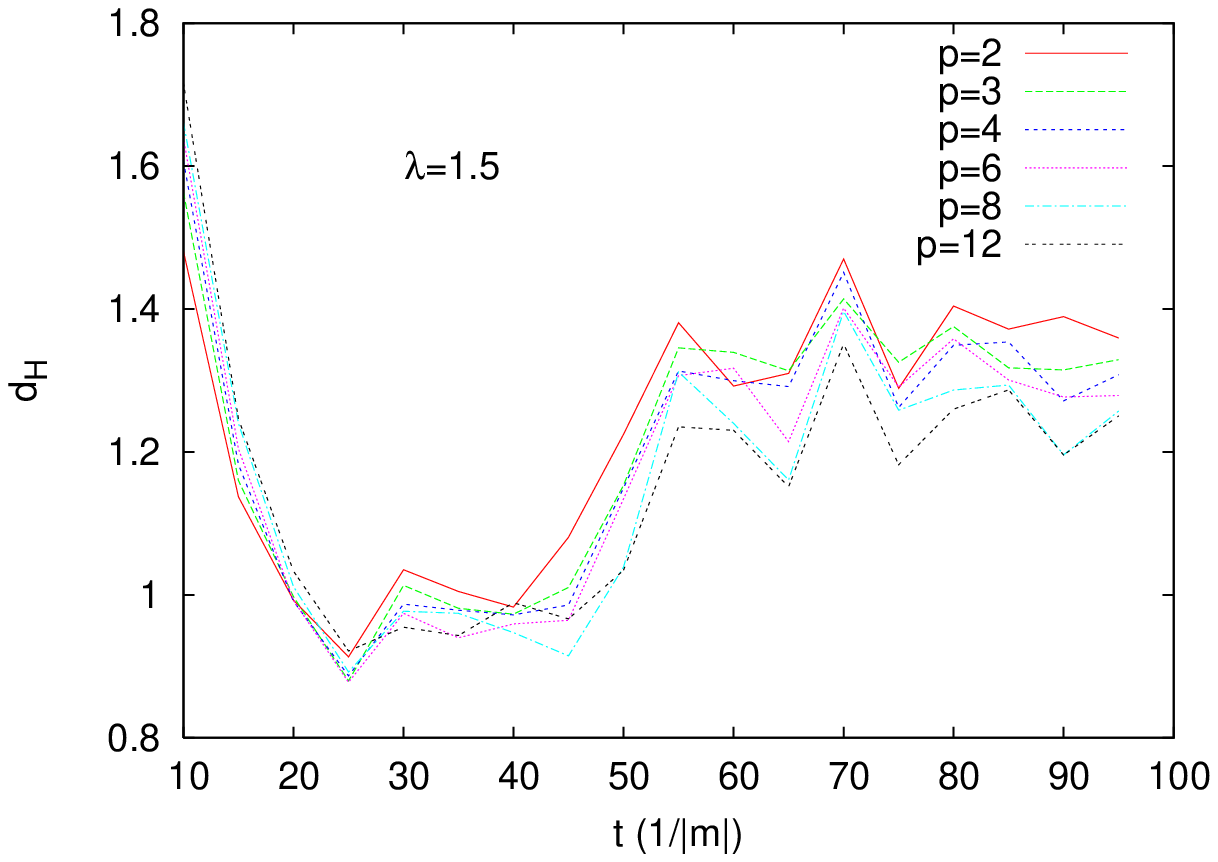}
\caption{ The time evolution of $d_H$ calculated from 
different $p$ values, typical for the distinct $\lambda$ regions (see text).  
}
\label{dHp}
}

 We have investigated first the time evolution of $d_H$ for solutions
belonging to the same value of $ |m|dx, \rho_{th}$, but different
$\lambda$. For each $\lambda$ value $n\geq 10$ independent runs
 were analysed, starting
from different initial conditions. For each run at a given time $|m|t$
blocking transformations with a number of scale factors
$p=2,3,4,6,8,12$ were
performed and $d_H(p)$ was calculated from (\ref{hausdorf}).  In
Figure \ref{dHp} we show the time variation of Hausdorff-dimensions
calculated with different $p$ for two characteristic values of
$\lambda$. In the left hand picture the unique curve, rather independent of
the actual value of $p$ proves that the functional form conjectured in
(\ref{hausdorf}) is obeyed. However, for such low values ($\lambda <0.5$)
the $d_H(t)$ function stays well above unity. We checked that in this
region no vortices winding around the lattice
stabilize for larger times, although the early temporary presence of
closed stringlike objects can be observed.  Very recently the production of
  vortices even with higher windings was demonstrated for $\lambda <
  3e^2$  \cite{donaire05}. The difference between the conclusions of
  the two studies could be twofold:
  i) our lattices  are 8-64 times smaller than those used in
  \cite{donaire05}, and these authors
 emphasized the need for very large size initial blobs of
  Higgs defects for vortex formation, ii) the difference between the
  dynamical processes leading to vortex formation in the two numerical 
simulations might be important.

 We had to exclude this $\lambda$-region ($\lambda<0.5$) from
the further analysis, because of the apparent absence of any time interval when
the system would be dominated by linear Higgs-defects. 

 The right hand curves exemplify the behavior of $d_H(t)$
  estimators from the use of different scale $p$ for the blocking in
  the range $\lambda >1$ where the unit value of $d_H(t)$ is reached
  at least for a
restricted time interval. In the present run the values arising
from using different values of $p$ show small dispersion until
$|m|t\leq 60$. The ''evaporation'' of the one-dimensional objects
apparently makes the concept of Hausdorff dimension less apropriate
for the Higgs-defect manifold.

\FIGURE{
\includegraphics*[width=15cm]{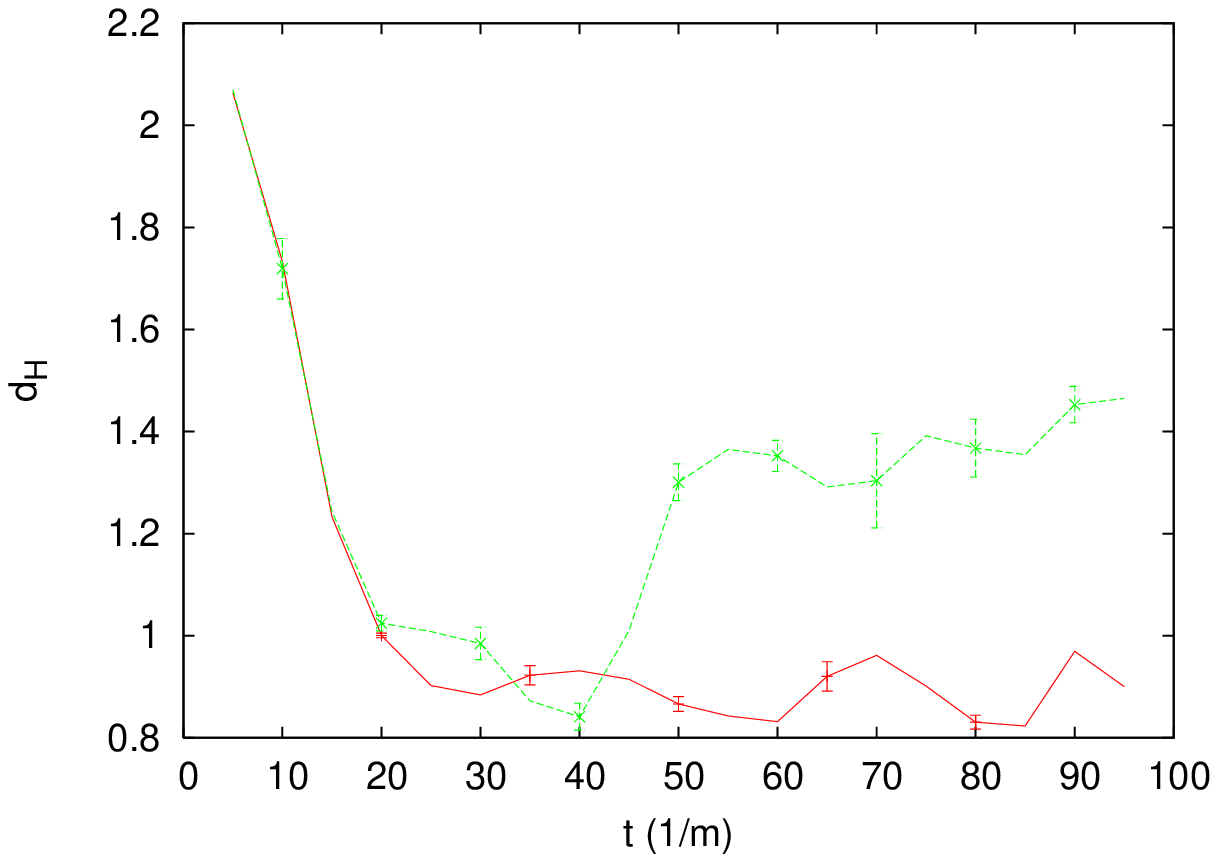} 
\caption{ Two typical shapes of the $d_H(t)$ function for $\lambda =3,
  |m|dx=0.35$ and $\rho_{th}/\rho_0=0.1$. Each point is an average over
$d_H(p)$ values obtained with 6 different blocking scales.  The
statistical error bars correspond to the spread of the $d_H$ values
calculated with different $p$. They 
do not depend on time when close to unity. 
The different shapes for $|m|t \geq 50$ sensitively
indicate the presence or absence of stable vortices for large times.}
\label{Fig:hausd_time}
}

For $\lambda\geq 1$  well-defined $d_H(t)$ functions emerge
 which consist of three
portions. The fast initial drop goes over into a nearly constant value
varying for different runs between $0.9$ and $1.1$. Then around
$|m|t=50$ in some runs the function changes steeply, but continously upwards 
and approaches
$d_H(|m|t\geq 100)\approx 1.4$. In other runs its fluctuations are
prolongated around the unit value. 
We checked that for the latter case stable vortices
winding around the lattice always appear. In the complementary case 
no macroscopic vortices get stabilized during the late time evolution,
 which is intuitively expected on the basis of the badly defined
 dimensionality of the Higgs zeros in this regime.
The two types of $d_H(t)$ functions are illustrated by the two curves 
 in Fig.\ref{Fig:hausd_time}.

In the allowed range of $\lambda$, $d_H$ was determined
in all runs in the interval $|m|t\in (20,40)$, where the effective
defect dimension was nearly one. We have computed the
volume fraction of the defects and their effective dimensionality 
as an average over this time interval. By this averaging the
  error bar of this estimate of the Hausdorff dimension was further
  decreased. The dependence of this time averaged value 
on the lattice spacing and  the value of $\rho_{th}$ was further investigated
as described next.

\FIGURE{
\includegraphics*[width=15cm]{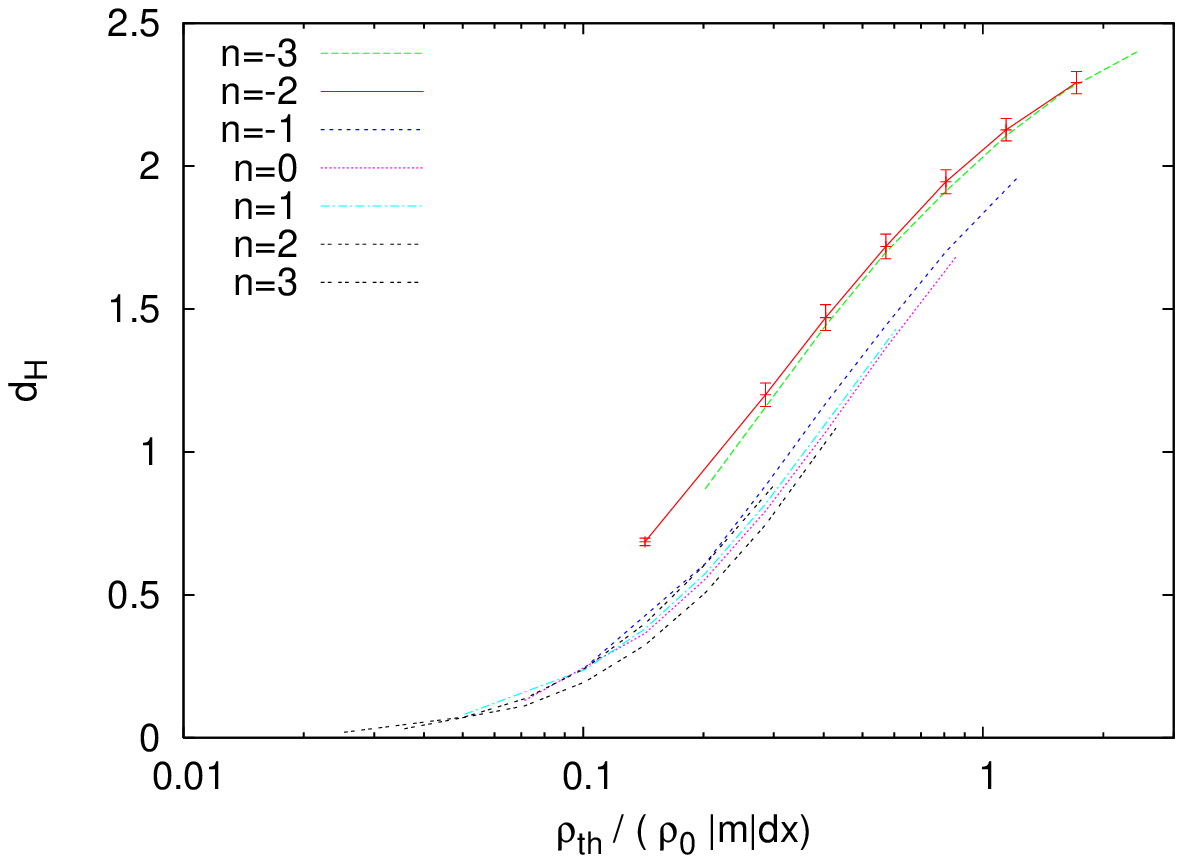} 
\caption{ The scaling variation of the Hausdorff dimension with
  $\rho_{th}$ for different values of $|m|dx$. $d_H$ represents the
  average of the values taken in the region $|m|t\in (20,40)$. 
The curves obtained with different $|m|dx$ should coincide 
if $d_H$ depends only on the
  ratio $\rho_{th}/(\rho_0|m|dx)$ and $\rho_{th}$ is scaled
  appropriately. The violation of the scaling for
  small $|m|dx$ signals nonlinear decrease of the Higgs field
  near the location of its zero. Typical error is shown on the $n=-2$ curve. }
\label{Fig:hausd_scale}
}
 
The next step of the analysis was the investigation of the 
dependence of $d_H$ on $|m|dx$, which is
 the other independent parameter of the discretized
equations of motion. A simple argument can be put forward which
suggests that (at least
for a certain range of $|m|dx$) its change can be compensated by the
variation of $\rho_{th}$, which governs the ''detection efficiency'' of 
vortices. In the interior of a vortex near its center
one can assume linear variation of the Higgs field with a slope
$\sim m_{Higgs}$. The expected value of the Higgs field at the nearest
lattice site can be estimated to be $\rho_0 m_{Higgs}dx$. Since
$m_{Higgs}\sim |m|$, therefore when $|m|dx$ changes,
the detectability of a vortex is kept on the same level only if the
''detection threshold'' $\rho_{th}$ is also varied accordingly:
\be
\rho_{th} \sim \rho_0|m|dx.
\ee 
 This observation predicts that the two-variable function
 $d_H(\rho_{th}/\rho_0,|m|dx)$ actually depends only on the ratio of the
 two. In Fig.\ref{Fig:hausd_scale} this scaling of $d_H$ is
 tested. The measurement was performed for dimensionless mass values
 characterized by the integer $n$:
 $|m|dx=0.35\times 2^{n/2}, n\in (-3,3)$.
The basic curve $(n=0)$ represents the variation of $d_H$ with 
$\rho^{(0)}_{th}=0.1$ for
 $|m|dx=0.35$.   It is obvious that the
 curves drawn with the shifted $\rho_{th}^{(n)}=2^{n/2} \rho_{th}^{(0)} ,
 n\in(-1,3)$ values coincide within the accuracy of the determination
 (cf. the error bars in Fig.\ref{Fig:hausd_scale}.).
  However, the shifted curves redrawn
 for the most negative values of $n$ (e.g. $-2,-3$) do not obey the scaling
 behavior, probably because the linear approximation with the slope
 $ m_{Higgs} $  is not valid very near to the zero of the Higgs field.
 The scaling $|m|dx$ dependence greatly simplifies our analysis, since
 it is sufficient to analyze the dependence of $d_H$ on $\lambda$ with
 a single $|m|dx$ and a conveniently chosen $\rho_{th}$.

It turns out that in the allowed range of $\lambda$
(e.g. $1\leq\lambda$) the effective
dimension gets  close to unity when using $ \rho_{th}=0.1\rho_0$. We fixed
eventually  those parameters ($|m|dx=0.35, \rho_{th}=0.1\rho_0$), 
whose variation is irrelevant for
investigating variations of the density of one-dimensional Higgs-defects. 
 It is the investigation of the dependence of
  the one-dimensional Higgs-defect density on the coupling $\lambda$,
  which remains our main task. 

The 
final step is to relate the proliferation of the vortexlike defects
to some typical time scale which characterizes the underlying
non-equilibrium phase transition. The usual Kibble--Zurek analysis
\cite{laguna97,laguna98} assumes the
existence of a second order transition and compares the relaxation
time of the system to the quenching time, which characterizes the
speed of the order parameter variation during the non-equilibrium phase
transition. The density of the defects is determined by the
correlation length calculated in the moment of the equality of the
relaxation and the quenching times, where the system falls out of
equilibrium. 

In systems of hybrid inflation the coupling of Higgs-fields to
the inflaton perfectly realizes this scenario near the critical
inflaton amplitude \cite{copeland02}. In our present investigation,
however, the system starts by the sudden quench
from an initial state far from equilibrium
and the sensitivity to  the nature of the
symmetry breaking (crossover, 1st or 2nd order transition) is not obvious.

We propose to introduce a new type of characteristic time, to be
called the  
roll-down time $\tau_r$ as the time needed for the 
average Higgs-field to reach its first maximum ($\rho_{max}\geq\rho_0$):
\be
\tau_r=\int_0^{\rho_{max}}d\bar\rho\left(\frac{d\bar\rho}{dt}\right)^{-1},
\qquad \bar\rho(t)=\frac{1}{V}\int d^3x \rho({\bf x},t).
\ee 
 The times characteristic for the phase transition and that of the
quench are not related in the present case (the latter is actually zero).
It is clear that $\tau_r$ is much shorter than the time necessary to
reach equilibrium. $\tau_r$ is in some way related to the
spatial correlation length taken at early times by the following argument. 
Since the spinodal (or ``tachyonic'') instability excites 
modes with wave numbers $k\leq |m|$, therefore directly after its saturation 
the typical size of homogenous domains is $\xi\sim |m|^{-1}$. 
The question is, how the roll-down time scales with $|m|$?

We have varied $\lambda$ by keeping the height of the potential
$V(0)-V(\rho_0)=3m^4/2\lambda$ fixed. 
In this way the  location of the minimum of 
the potential was varied
$\rho_0=(-6m^2/\lambda)^{1/2}\sim |m| ^{-1}$. Taking into 
account the fixed height of the potential also 
$\tau_r\sim \lambda^{-1/4}$ is valid.
Therefore the smaller is $\lambda$ the longer is the
transition time $\tau_r$. We have measured
its dependence on $\lambda$
through the vacuum expectation values of the Higgs-field: 
\be
  \tau_r \sim \rho_0^{1/y}\sim |m|^{-1/y}, \qquad 1/y=0.64 \pm 0.01. 
\ee
This relation reveals a space-time anisotropy in this sytem, the
roll-down time and the spatial correlation length scale differently.
It leads to a nontrivial mapping of the $\lambda$-dependence of the
defect density into a function of $\tau_r$.

\FIGURE{
\includegraphics*[width=15cm]{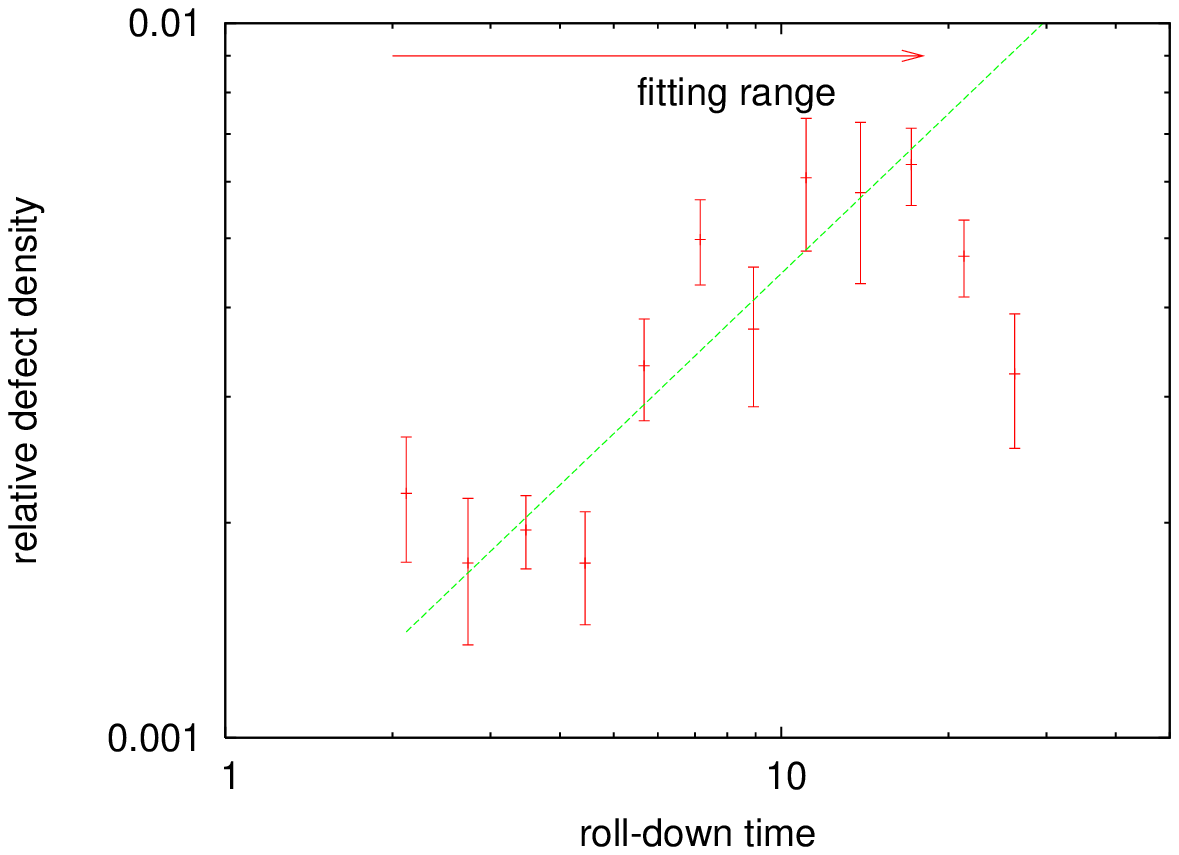} 
\caption{ The variation of the volume fraction of the Higgs defects
  defined at $\rho_{th}=0.1 \rho_0$ with the roll-down time. On log-log
  scale Kibble-Zurek scaling implies variation along a straight line.
}
\label{Fig:KZ}
}

The defect density was    fitted with an exponential decay in the time 
range: $(20,40)|m|^{-1}$. 
The fit was then extrapolated back to time $\tau_r$ 
to get the initial defect density $N_0$ .  The dependence of 
the initial defect density on $\tau_r$ is displayed in
Fig.\ref{Fig:KZ}. In the range displayed on the figure a power law fit
was made to it:
\be
  N_0 \sim \tau_r ^z,\qquad  z= 0.7 \pm 0.4 .
\ee
The large error is the sum of two effets. The statistical error
  of the slope determination contributes cca. $0.15$, the rest is the
  result of some systematics related to the choice of the time
  interval where the exponential fit is made.

If we are interested in the density of
the vortices we have to count only once the Higgs defect sites
belonging to the same vortex, that is we have to divide $N_0$ by the
average length of a vortex, e.g. $\xi$.  Therefore
we find
\be
n_{vortex}\sim \tau_r^{z-y}, \qquad z-y\approx -0.9\pm 0.4.
\ee
This estimate hints at a sharper decrease of the vortex density than
 proposed in
\cite{laguna98}, but the large error does not allow to draw a definite
conclusion yet whether the vortex density decreases with an exponent different
from the case discussed in \cite{kibble76,zurek85}.

\section{ Conclusions}

We have presented a detailed study of the temporal
evolution during the non-equilibrium transition leading to the
Higgs-effect in classical scalar electrodynamics at sufficiently low
energy density. The quick appearance of the degeneracy of longitudinal
and transversal gauge excitations is reflected both in the straight
line trajectory of these degrees of freedom in the $(p -
\epsilon)$-plane, and by the
spectral equations of state except the modes with the 
lowest and the very high wave numbers. This represents a clear
evidence for the so-called prethermalisation \cite{borsanyi04}.
On the other hand the degree of
excitation of the longitudinal modes is much higher in tachyonic
instability, and they relax very slowly. They might decouple during the
cosmological expansion with a higher
effective temperature than the transversal modes do. Late time 
decay of the hotter longitudinal modes 
might produce more energetic charged particles which could 
participate in elementary processes
of cosmological interest.

In this paper also a method was proposed for
a refined estimate of the density of stringlike objects present in the whole
sample for a certain range of the coupling $\lambda$. The change in
the other coupling $|m|dx$ always could be compensated by the
resolution parameter $\rho_{th}$.
The measurement was performed and averaged in an intermediate time
interval ($|m|t\in (20,40)$) 
and the resulting defect density displays powerlike behaviour as a
function of the time characteristic for the transition from the
unstable into the stable symmetry breaking minimum.  The accuracy of
the determination of its power at present is rather poor. 
Our result indicates that the Kibble--Zurek-scaling phenomenon 
(not neceesarily with a universal scaling power)
may not be tied to the existence of an equilibrium
second order transition in the system.

\section*{Acknowledgement}
The authors benefited from the very valuable comments of
M. Hindmarsh. We thank Z. Haiman for a discussion on possible astrophysical
implications. 
This research was supported by the Hungarian Research Fund under the
contract No. T046129.

\end{document}